\documentclass[twocolumn]{article}

\usepackage{authblk}

\usepackage[hidelinks]{hyperref}
\usepackage[linesnumbered,ruled]{algorithm2e}
\usepackage[most]{tcolorbox}
\usepackage[small]{caption}
\usepackage[switch]{lineno}
\usepackage[utf8]{inputenc}
\usepackage{amsmath}
\usepackage{amssymb}
\usepackage{amsthm}
\usepackage{booktabs}
\usepackage{enumitem}
\usepackage{fancyvrb}
\usepackage{fvextra}
\usepackage{graphicx}
\usepackage{longtable}
\usepackage{mathtools}
\usepackage{microtype}
\usepackage{multirow}
\usepackage{natbib}
\usepackage{numprint}
\usepackage{siunitx}
\usepackage{soul}
\usepackage{times}
\usepackage{url}
\usepackage{verbatim}
\usepackage{wrapfig}

\tcbset{
  observationbox/.style={
    colback=blue!5,
    coltitle=white,
    coltext=black,
    boxrule=0.2mm,
    sharp corners,
    title={Observation},
    before skip=5pt,
    after skip=5pt,
    enhanced,
  }
}

\newcommand{\ccfgtf}[1][]{{\textsc{CcfgT5}\textsubscript{{#1}}}}

\newcommand{\sectref}[1]{{Section~\ref{#1}}}

\newcommand{\exampleref}[1]{{Example~\ref{#1}}}
\newcommand{\tableref}[1]{{Table~\ref{#1}}}

\newtheorem{example}{Example}

\title{
    LogiCase: Effective Test Case Generation from Logical Description 
    in Competitive Programming
}

\author[1]{Sicheol Sung}
\author[2]{Aditi}
\author[3]{Dogyu Kim}
\author[1]{Yo-Sub Han}
\author[2]{Sang-Ki Ko\thanks{Corresponding author: sangkiko@uos.ac.kr}}

\affil[1]{Yonsei University, \texttt{\{sicheol.sung, emmous\}@yonsei.ac.kr}}
\affil[2]{University of Seoul, \texttt{\{aditimzu16, sangkiko\}@uos.ac.kr}}
\affil[3]{Kangwon National University, \texttt{dogyu.kim9@gmail.com}}

\date{}

\begin{document}

\maketitle

\begin{abstract}
Automated Test Case Generation~(ATCG) is crucial for evaluating software
reliability, particularly in competitive programming where robust algorithm
assessments depend on diverse and accurate test cases. However, existing ATCG
methods often fail to meet complex specifications or generate effective corner
cases, limiting their utility. In this work, we introduce Context-Free Grammars
with Counters~(CCFGs), a formalism that captures both syntactic and semantic
structures in input specifications. Using a fine-tuned CodeT5 model, we
translate natural language input specifications into CCFGs, enabling the
systematic generation of high-quality test cases. Experiments on the
CodeContests dataset demonstrate that CCFG-based test cases outperform baseline
methods in identifying incorrect algorithms, achieving significant gains in
validity and effectiveness. Our approach provides a scalable and reliable
grammar-driven framework for enhancing automated competitive programming
evaluations.
\end{abstract}

\begin{figure*}[ht!]
\centering
\includegraphics[width=.8\textwidth,page=1]{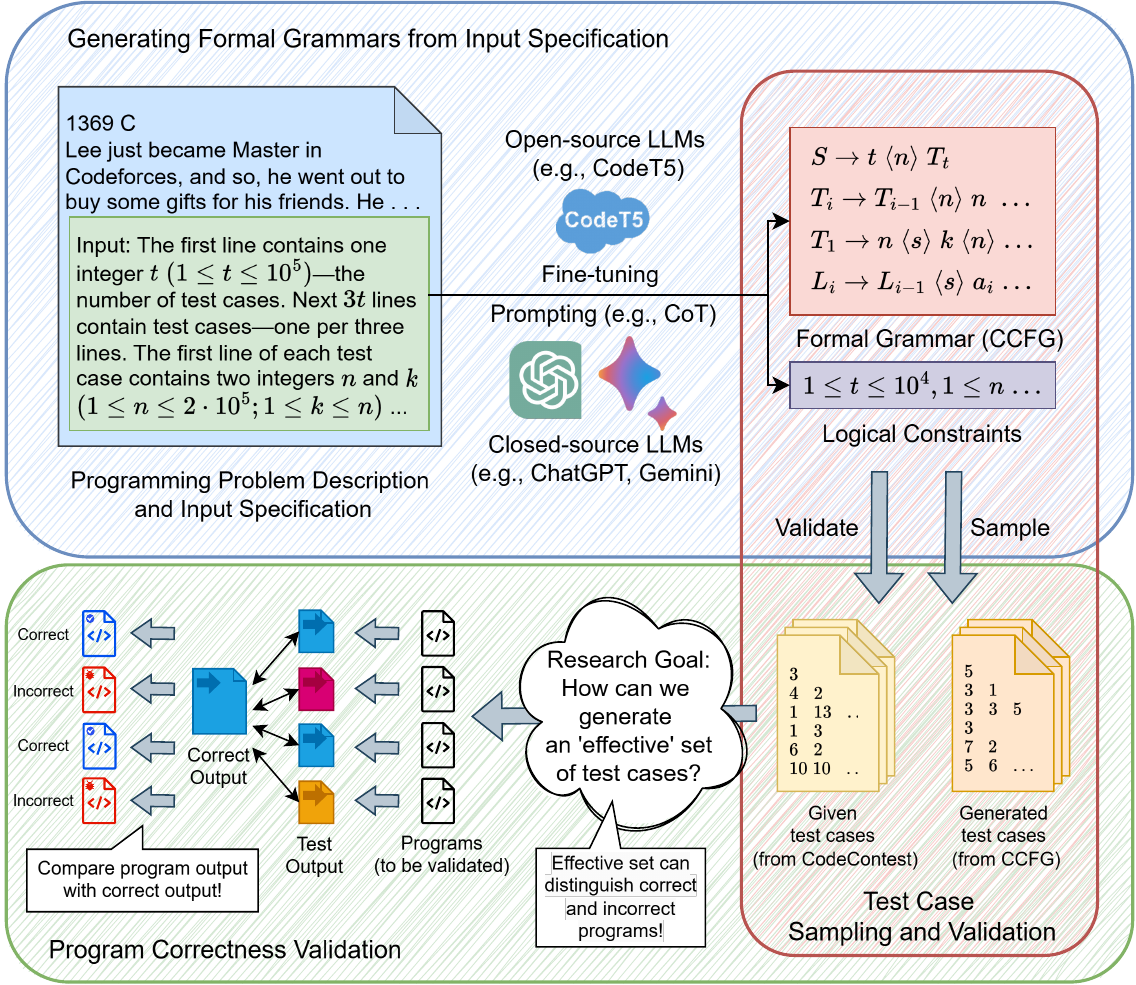}
\caption{
    Overview of the proposed framework for generating test cases for competitive
    programming problems. The deep learning model translates specifications into
    CCFGs while preserving their meaning. Subsequently, the CCFGs are utilized
    to generate test cases.
}
\label{fig:overview}
\end{figure*}

\section{Introduction}

Automated Test Case Generation~(ATCG)~\citep{AnandBCCCGHHMOE13} is a critical
component of software engineering, particularly in competitive programming,
where the performance and correctness of algorithms are tested against diverse
and challenging inputs. As the complexity of software and algorithmic problems
increases, manual test case creation becomes infeasible, prompting the rise of
automated solutions. Recent advancements in deep learning have significantly
improved the generation of test cases, yet several fundamental challenges
persist. Ensuring adherence to intricate input specifications and capturing
corner cases effectively remain elusive, often necessitating additional manual
analysis.

The inadequacy of test suites has been a persistent issue in software testing.
For example, the Defects4J dataset, as highlighted by~\cite{FraserA11}, has
shown that incomplete test cases lead to insufficient program analysis.
Similarly, the CodeNet dataset, a benchmark for competitive
programming~\citep{Puri0JZDZD0CDTB21}, suffers from a lack of high-quality test
cases, as noted by~\cite{ZhaoHMLZJLZ024}. These limitations not only hinder
robust algorithm validation but also exacerbate challenges in program
repair~\citep{TianLPKLHKB22}. Even recent prompt-based methods leveraging Large
Language Models~(LLMs), such as ChatGPT, and mutation-based frameworks like
MuTAP~\citep{DakhelNMKD24}, often produce invalid or incomplete test cases for
competitive programming due to the complexity of test case specifications. 

These shortcomings highlight a critical gap in automated testing: existing
methods struggle to generate test cases that adhere strictly to complex problem
constraints while capturing edge cases essential for algorithm validation. For
instance, inaccuracies in some program synthesis benchmarks, as revealed
by~\cite{LiuXW023}, further emphasize the necessity of robust testing frameworks
capable of addressing these gaps.

To tackle these challenges, we propose Context-Free Grammars with
Counters~(CCFGs), a novel formalism that integrates both syntactic and semantic
elements of problem input specifications. By leveraging a fine-tuned
CodeT5~\citep{WangWJH21} model, our approach translates natural language
descriptions into CCFGs, enabling the automated generation of test cases that
are valid, specification-compliant, and highly effective in uncovering
algorithmic flaws. Figure~\ref{fig:overview} illustrates our proposed
framework.

Our main contributions are threefold:
\begin{itemize}
    \item We introduce CCFGs to encode problem input specifications, capturing
    both syntax and semantics in a unified framework tailored for competitive
    programming.
    \item We develop a specialized model to map natural language problem
    descriptions to CCFGs, ensuring precise and specification-compliant test
    case generation.
    \item Using the CodeContests dataset~\citep{LiAlphaCode22}, we demonstrate
    that CCFG-based test cases significantly outperform baseline methods in
    identifying algorithmic errors and distinguishing correct from incorrect
    solutions.
\end{itemize}

Note that the CodeContests dataset is a powerful resource for benchmarking and
evaluating test case generation methods due to its diverse collection of
competitive programming platforms such as AtCoder, CodeChef, Codeforces, etc.

One of the key strengths of our approach lies in its ability to overcome the
primary limitation of grammar-based test case generation techniques---the need
for a manually written input grammar for each specification. Unlike traditional
methods that require crafting a dedicated grammar for every input specification,
our approach generalizes seamlessly across all input specifications in
competitive programming problems. This eliminates the need for extensive manual
effort and expertise to define grammars, enabling broader applicability and
reducing the risk of incomplete or erroneous test coverage. By bypassing the
reliance on handcrafted grammars, our method ensures scalability and efficiency,
adapting to diverse input formats without requiring tailored adjustments for
each new specification. This capability not only simplifies the testing process
but also enhances its robustness and versatility, making it particularly suited
for dynamic and varied environments like competitive programming.

\section{Related Work}

\subsection{Test Case Generation using LLMs}

Recent advancements in LLMs have opened new possibilities for ATCG by leveraging
their ability to understand and generate structured data, natural language, and
code with remarkable accuracy~\citep{FengGTDFGS0LJZ20, GuoLDWZY22,
LiAlphaCode22, RoziereGG23,MesnardHD24}.

TestAug~\citep{YangHS0022} and ChatTESTER~\citep{YuanLLDWCP23} utilized ChatGPT
to derive test cases directly from natural language descriptions of code, often
accompanied by code snippets. These methods provide a straightforward way to
automate test case generation but face limitations in covering complex program
logic and edge cases due to the inherent difficulty in aligning natural language
prompts with precise program semantics. TestEval~\citep{WangYWHCSZCM24}
introduced a dataset of 210 Python programs sourced from LeetCode and designed
three evaluation tasks: overall coverage, targeted line/branch coverage, and
targeted path coverage.

CodaMosa~\citep{LemieuxILS23} utilizes a programming dataset to generate test
cases by leveraging the LLM (OpenAI's Codex~\citep{Codex21}). Codex produces
initial test cases that are syntactically and semantically meaningful, providing
a strong foundation for further exploration. These test cases are then
iteratively mutated to maximize code coverage. However, while CodaMosa
effectively optimizes for code coverage, this metric does not always correlate
with fault detection or real-world program behavior.

Recently, \cite{XiaPTP024}, introduced Fuzz4All generating test cases of many
different languages by leveraging the multi-lingual capabilities of LLMs. This
approach accepts inputs in various formats, such as example code snippets,
program specifications, or documentation, and uses multiple prompts to guide the
LLM in generating fuzzing inputs. 

\subsection{Grammar-based Test Case Generation}

Grammar-based fuzzing is well-established in software testing that uses formal
grammars to guide the generation of inputs. The foundational work on fuzz
testing by \citet{MillerFS90} laid the groundwork for input generation
techniques, demonstrating the effectiveness of random input generation. However,
random fuzzing techniques often fail when targeting structured inputs because
they are unlikely to produce valid test cases. On the other hand, grammar-based
fuzzing, which relies on CFGs to generate syntactically valid test inputs,
marked a substantial improvement over purely random input generation techniques
by enabling more targeted testing. 

Grammar-based white-box fuzzing~\citep{GodefroidKL08} enhances traditional
grammar-based fuzzing by incorporating symbolic execution, enabling the
exploration of program paths. \citet{SrivastavaP21} have proposed Gramatron,
which uses {\em grammar automatons} in conjunction with aggressive mutation
operators to synthesize bug triggers faster. 

Note that a significant limitation of grammar-based test case generation or
fuzzing techniques is the requirement for a manually written input grammar. This
dependency introduces several challenges, particularly in terms of time,
expertise, and coverage. Crafting a comprehensive input grammar requires a deep
understanding of the input structure, including syntax and semantics, which
demands significant effort and domain-specific knowledge from the developer or
tester.

\section{Methodology}

We introduce the Context-Free Grammars With Counters~(CCFG) to mitigate the
limitations of previous approaches to automated testing of competitive
programming. CCFGs are tailor-made to capture the formal semantics of most
programming competition problems in the concise form of formal grammar by
utilizing the counters to capture numerical values specified in the
specifications.

\subsection{Context-Free Grammars With Counters}
\label{sec:ccfg}

Let us start with a motivating example. Example~\ref{example:specification} is
an input specification of ``1369\_C. RationalLee'' from
Codeforces.\footnote{\url{https://codeforces.com/problemset/problem/1369/C}}

\begin{example}[Input Specification]\label{example:specification}
\hfill
\begin{itemize}[leftmargin=1.0em,itemsep=-0.2em]
    \item The first line contains one integer~$t$.
    \item Next $3t$~lines contain test cases---one per three lines.
    \item The first line of each test case contains two integers~$n$ and~$k$.
    \item
    The second line of each test case contains $n$~integers~$a_1, \ldots, a_n$.
    \item The third line contains $k$~integers $b_1, \ldots, b_k$.
\end{itemize}
\end{example}

To explain the necessity of CCFGs, let us describe the input specification of
Example~$\ref{example:specification}$ using a CFG as follows; for simplicity, we
assume we modify a generation algorithm of CFG to sample values of
variables~$t$, $n$, $k$, $a_i$ and $b_i$ according to their constraints during
the generation.

\begin{example}[\underline{Incorrect} CFG of Example~\ref{example:specification}]
\label{example:context_free_grammar}
\begin{align*}
    S &\to t\ \texttt{<n>}\ T, \\
    T &\to T\ \texttt{<n>}\ n\ \texttt{<s>}\ k\ \texttt{<n>}\ L\ \texttt{<n>}\ Z \\
    &\hspace{6pt} \vert \hspace{7pt}
        n\ \texttt{<s>}\ k\ \texttt{<n>}\ L\ \texttt{<n>}\ Z, \\
    L &\to L\ \texttt{<s>}\ a_i \mid a_i, \\
    Z &\to Z\ \texttt{<s>}\ b_i \mid b_i.
\end{align*}
\end{example}

While the above grammar can generate all valid test cases, it also generates
invalid ones. During generation, each application of the production rule for~$T$
produces a single test case. Therefore, we need to limit the number of such
applications to~$t$, where the value of~$t$ is determined during the generation.

Intuitively, CCFG restricts the number of applications of production rules by
using a {\em counter} to track the number of applications and selecting the next
production rule based on the counter's value. Using this approach, we can modify
the CFG of Example~\ref{example:context_free_grammar} to obtain the correct CCFG
in Example~\ref{example:counting_context_free_grammar}.

\begin{example}[\underline{Correct} CCFG of Example~\ref{example:specification}]
\label{example:counting_context_free_grammar}
\begin{align*}
    S &\to t\ \texttt{<n>}\ T_t, \\
    T_i &\to T_{i-1}\ \texttt{<n>}\ n\ \texttt{<s>}\ k\ \texttt{<n>}\ L_n\
    \texttt{<n>}\ Z_k, \\
    T_1 &\to n\ \texttt{<s>}\ k\ \texttt{<n>}\ L_n\ \texttt{<n>}\ Z_k, \\
    L_i &\to L_{i-1}\ \texttt{<s>}\ a_i,\quad L_1 \to a_1, \\
    Z_i &\to Z_{i-1}\ \texttt{<s>}\ b_i,\quad Z_1 \to b_1.
\end{align*}
\end{example}

Next, we generate test cases using the CCFG as follows. When applying the
production rule for $T_i$ to $T_t$, we set the value of an internal counter to
$t$. Then, during the derivation of $T_{i-1}$, we decrement the counter by 1.
Finally, when the counter value reaches $1$, we apply the production for $T_1$
to $T_i$. 

Note that the proposed CCFG is substantially different from the grammar studied
by~\cite{ChistikovHH18} as our CCFG can associate the integer input with
non-terminals by pre-defined production rules during the parsing while Chistikov
{\em et al.} consider a grammar that resets or adds integers to a pre-defined
set of counters at each step.

\subsection{CCFGT5 Translation Model}

Inspired by the idea of {\em grammar prompting}~\citep{WangW0CSK23}, which
enables LLMs to use domain-specific constraints expressed through a formal
grammar, we propose a translation model, \ccfgtf, designed to translate a
problem input specification in NL into a concise CCFG while preserving the
original semantics. This model incorporates two specifically fine-tuned CodeT5
modules: one focuses on grammar, and the other on constraints. We use Adam
optimizer with learning rate~$10^{-5}$ and cross-entropy loss function to train
each CodeT5 model. We generate candidate grammars and constraints with
repetition penalty~$2.5$ and length penalty~$1.0$ from each model.

We also use a specialized CCFG tokenizer to enable effective training. Our
tokenizer converts a CCFG into a list of grammar symbols and labels each symbol
type with descriptive words, such as `variable' and `nonterminal,' to enhance
the readability from the LLM's perspective.

\section{Experiments}

We evaluate the practical usefulness of CCFGs through experimental validation.
All implementations and associated codes and datasets used in these experiments
are available in our GitHub
repository.\footnote{\url{https://github.com/Aditi1612/Grammar-based-test-case-generation}}

\subsection{Dataset}
\label{sec:dataset}

We use the CodeContests dataset, which consists of various programming problems
sourced from different competitive platforms~\citep{LiAlphaCode22}. This dataset
includes algorithms for programs in various programming languages and
\emph{public}, \emph{private}, and \emph{generated} test cases for each problem.

We manually created CCFGs for 1,500 different problems based on their
descriptions and categorized the human-labeled grammars into three levels:
\emph{easy}, \emph{normal} and \emph{hard}. Easy problems consist of simple
grammars that consist solely of variables, without any complex structures or
additional elements. Normal problems include grammars with one nonterminal that
requires counting or tracking. Lastly, hard problems encompass grammars that
have more than one nonterminal. These grammars are the most complex, involving
multiple non-terminal elements and adding layers of complexity and structure.
After categorizing the grammars, we split them into a training dataset with
1,200~problems and an evaluation dataset with 300~problems. Notably, the
training dataset contains more difficult grammars to help the model learn
complex syntactic structures effectively. \tableref{tab:dataset} summarizes the
difficulty distribution of each dataset.

\begin{table}
\caption{
    Distribution of difficulty in the dataset. We exclude 29 problems from the
    test dataset that (1)~lack incorrect solutions or (2)~our implementation
    cannot process their human-labeled grammars.
}
\label{tab:dataset}
\centering







\begin{tabular}{lrrrr}
\toprule

\multirow{2}{*}[-2pt]{\textbf{Category}} &
\multicolumn{3}{r}{\textbf{Specification difficulty} (\#)} &
\multirow{2}{*}[-2pt]{\textbf{Total}} \\

\cmidrule{2-4}

&
Easy & Normal & Hard
& \\

\midrule

Train & 518 & 553 & 129 & 1,200 \\
Evaluation & 159 & 101 & 11 & 271 \\

\bottomrule
\end{tabular}
\end{table}

We filter out 6~problems from the evaluation dataset for which CodeContests has
no incorrect solutions that are necessary for evaluation. Additionally, we
exclude 23 more problems that our CCFG implementation cannot fully support, due
to the combinatorial complexity of input specification.

\subsection{Test Case Sampling}

As in Example~\ref{example:specification}, the value of a variable in a test
case often determines the number of subsequent lines or variables in that test
case. Therefore, we can control the length of the generated test cases by
varying the interval of variable sampling during the generation. Instead of
using the original interval~$(n, n+k)$, we sample a value of a variable from one
of the following options: (1)~the interval~$(n, n+k)$, (2)~the interval~$(n, n +
\log k)$, (3)~$(n, n+\log\log k)$, or (4)~the minimum value~$n$. When generating
ten \emph{long} test cases, we first create a test case using option~(1). If it
fails (e.g., due to the timeout), we then try the options~(2) and (3) in order.
We also generate ten~\emph{medium} and \emph{short} test cases by starting with
options~(2) and~(3), respectively. Finally, if we succeed in generating a corner
case with option~(4), we replace one of the short test cases with the corner
case.

\subsection{Baselines}

\paragraph{Mutation-based fuzzing.}

We utilize the public and private test cases from the CodeContests dataset,
tokenizing these test cases based on spaces and newline characters. We then
randomly select 30\% of tokens for mutation, adapting our approach according to
the token type: integer, float, or string. This selective mutation process
enables effective fuzzing while still adhering to the original input
specifications. 

\paragraph{Direct test case generation from LLMs.}

We employ two LLMs, OpenAI's ChatGPT 4 and Google's Gemini, to generate test
cases directly. We provide an input specification and a strict format required
for generating the test cases. To fully exploit the performance of LLMs, we use
the Chain-of-Thought~(CoT)~\citep{Wei0SBIXCLZ22} style prompt.

\begin{table*}[t!]
\caption{
    Validity and effectiveness of the test cases across different methods.
    \ccfgtf[n] refers the grammars generated by \ccfgtf~model with
    beam-size~$n$. We select a pair of grammar and constraints among top-$k$
    grammars and constraints. Gemini-$n$ and ChatGPT-$n$ refer to the CCFGs
    produced by LLMs employing CoT with $n$~different examples. Note that
    set-based effectiveness for both the CodeContests and Fuzzing categories,
    marked with an asterisk~(*), may involve more or fewer than 10~test cases for
    each problem, which limits the comparability of these results with the
    set-based effectiveness from other methods.
}
\centering
\label{tab:validity_effectiveness}
\sisetup{round-mode=places,round-precision=2,detect-weight=true}
\begin{tabular}{ll@{}cSSSS}
\toprule

\multirow{2}{*}[-2pt]{\textbf{Category}} &
\multirow{2}{*}[-2pt]{\textbf{Method}} &
\multirow{2}{*}[-2pt]{\textbf{Well-defined} (\#)} &
\multicolumn{2}{c}{\textbf{Validity} (\%)} &
\multicolumn{2}{c}{\textbf{Effectiveness} (\%)} \\

\cmidrule(lr){4-5}
\cmidrule(lr){6-7}

& & &
{Element-based} & {Set-based} &
{Element-based} & {Set-based} \\

\midrule[\heavyrulewidth]

&
Public &
270 & 99.63099631 & 99.63099631 & 31.70984888 & 39.6340713407134{$^*$} \\

CodeContests &
Private &
208 & 76.75276753 & 76.75276753 & 39.93422773 & 70.8948339483394{$^*$} \\

&
Generated &
269 & 77.9544878 & 30.25830258 & 18.25304912 & 28.0811808118081{$^*$} \\

\midrule
\midrule

\multirow{2}{*}{Direct gen.} &
Gemini &
271 & 80.1476014760147 & 61.2546125461254 & 26.6219762197622 & 44.1758917589175 \\
&
ChatGPT &
271 & \bfseries 91.38376384 & 78.96678967 & 37.7475839 & 63.53672465 \\
\midrule

\multirow{2}{*}{Fuzzing} &
Public &
270 & 79.26199262 & 45.01845018 & 17.4030926 & 28.7717448603057{$^*$} \\

&
Private &
208 & 63.65313653 & 33.57933579 & 17.58689158 & 28.1567387102442{$^*$} \\

\midrule

\multirow{7}{*}{CCFG-based} &
$\text{Gemini}_1$ &
108 & 16.05761219 & 15.49815498 & 8.06213546 & 13.13653137 \\
&
$\text{Gemini}_5$ &
158 & 46.04213784 & 44.64944649 & 24.01618855 & 39.74169742 \\

\addlinespace[0.2em]

&
$\text{ChatGPT}_1$ &
211 & 67.14676824 & 66.42066421 & 32.15549736 & 54.04366544 \\
&
$\text{ChatGPT}_5$ &
226 & 78.43113915 & 77.8597786 & 37.81206602 & 65.28597786 \\

\addlinespace[0.2em]

&
$\textsc{CcfgT5}_1$ &
132 & 19.56909892 & 18.81918819 & 10.08133952 & 15.90405904 \\
&    
$\textsc{CcfgT5}_{10}$ &
264 & 82.32353291 & \bfseries 81.18081181 & \bfseries 42.26064357 & \bfseries 67.72755228 \\

\midrule
\midrule

CCFG-based & Ground-truth &
271 & 100 & 100 & 52.51200254 & 83.40405904 \\
\bottomrule
\end{tabular}
\end{table*}

\subsection{Evaluation Metrics}\label{sec:metrics}

We carefully design evaluation metrics to address the following research
questions throughout experimental results: (1)~Is the CCFG-based approach better
than direct test case generation? (2)~Which method is most effective for
generating CCFGs from descriptions?

\paragraph{Validity and generality.}

We say that a test case is {\em valid} if it follows the logical input
specification of the problem. We evaluate the validity of each test case by
determining whether or not the ground-truth grammar can parse the test case.
Since it is computationally undecidable~\citep{HopcroftMU2001} to decide whether
or not a given CCFG is valid, we instead empirically measure \emph{element-based
validity} of grammar as the ratio of valid test cases to the total number of
test cases. Additionally, we measure \emph{set-based validity} of a grammar by
checking whether all generated test cases are valid. The value of the set-based
validity is either 0 or 1, while element-based validity can take any value in
between. We say a set of test cases is \emph{valid} if its set-based validity
is~1. 

Note that the validity alone cannot ensure that the grammar generates all the
possible test cases described by the input specification. To measure how many
valid test cases can be covered by a grammar, we define \emph{element-based
generality} of a grammar as the ratio of test cases that can be parsed by the
grammar to the total number of test cases generated by the ground-truth grammar.
A \emph{set-based generality} of a grammar is $1$ if and only if the test
case-based generality is also $1$. In this case, we call the grammar is
\emph{general}; otherwise, the set-based generality is $0$. If the set of test
cases generated by a grammar is valid and the grammar is general, we say that
the grammar is (empirically) \emph{semantically equivalent} to the ground-truth
grammar. In contrast, we say that a grammar is {\em syntactically equivalent} to
the ground truth if two grammars are equivalent except for the naming of
variables.

\paragraph{Effectiveness.}

The primary purpose of test cases in competitive programming is to distinguish
between correct and incorrect algorithms. For a given problem~$p$, let $A_p$ be
a set of all incorrect algorithms implying that for each $y \in A_p$, there
always exists a valid test case $x$ such that $\hat{y}(x) \ne y(x)$, where
$\hat{y}(x)$ is the correct output for $x$. Then, we define the
\emph{effectiveness}~$E(x, A_p)$ of a test case~$x$ with respect to~$A_p$ as
\[
    E(x, A_p)
    := \frac{|\{y \in A_p \mid y(x) \neq \hat{y}(x)\}|}{|A_p|},
\]
which is the ratio of incorrect algorithms in $A_p$ that are {\em
distinguishable} by the test case $x$ to all incorrect algorithms $A_p$. We
extend to define the \emph{element-based effectiveness}~$E_\text{elt}(X, A_p)$
of a set~$X$ of test cases with respect to $A_p$ as
\[
    E_{\text{elt}}(X, A_p) :=
    \begin{cases}
        \frac{1}{|X|} \sum_{\substack{x \in X}} E(x, A_p), &
        \text{if $X$ is valid}, \\
        \hfil 0, & \text{otherwise}.
    \end{cases}
\]

Note that the element-based effectiveness of a set of test cases is the average
effectiveness of individual test cases in the set. If each test case can
distinguish different incorrect algorithms, then the entire set can identify
more incorrect algorithms compared to any single test case. Therefore, the
diversity of the algorithms that the set can distinguish is also important.
Thus, we define \emph{set-based effectiveness}~$E_\text{set}(X, A_p)$ of a
set~$X$ of test cases as
\begin{align*}
    E_{\text{set}}(X, A_p) & \\
    := \begin{dcases}
        \frac{
            |\{y \in A_p \mid y(x) \neq \hat{y}(x), \exists x \in X\}|
        }{|A_p|}, &
        \text{if $X$ is valid}, \\
        \hfil 0, & \text{otherwise}.
    \end{dcases} \span
\end{align*}

For experiments, we determine the correct output~$\hat{y}(x)$ for a test
case~$x$ by executing up to ten correct algorithms from the dataset and
selecting the most frequently occurring output as the correct one. Additionally,
for each problem~$p$, we sample at most ten incorrect algorithms to create a set
designated as~$A_p$. We treat the output of algorithms that exceed twice the
original timeout as~$\bot$, which is always considered incorrect.

We use a total of ten test cases from thirty test cases generated by CCFGs to
compute set-based effectiveness, consisting of four short test cases (including
corner case if possible) and three medium and long test cases. This approach
ensures that we are using the same number of test cases as in the case of direct
generation.

\begin{table*}[htbp]
\caption{
    Validity and effectiveness with respect to the difficulty of input
    specifications for problems. The first two rows present results for the
    direct test case generation approach, while the remaining rows are for the
    CCFG-based approach.
}
\centering
\label{tab:difficulty}
\sisetup{round-mode=places,round-precision=2,detect-weight=true}
\begin{tabular}{
l
ccc
SSS
SSS
}
\toprule
\multirow{2}{*}[-2pt]{\bf Method} &
\multicolumn{3}{c}{\textbf{Well-defined} (\#)} &
\multicolumn{3}{c}{\textbf{Set-based validity} (\%)} &
\multicolumn{3}{c}{\textbf{Set-based effectiveness} (\%)}
 \\
\cmidrule(lr){2-4}
\cmidrule(lr){5-7}
\cmidrule(lr){8-10}
&
{Easy} & {Normal} & {Hard} &
{Easy} & {Normal} & {Hard} &
{Easy} & {Normal} & {Hard} \\

\midrule[\heavyrulewidth]

Gemini$_\text{(Direct)}$ &
159 & 101 & 11 &
71.0691823899371 & 47.5247524752475 & 45.4545454545454 &
53.4905660377358 & 31.5511551155115 & 25.4545454545454 \\

ChatGPT$_\text{(Direct)}$ &
159 & 101 & 11 &
\bfseries 88.67924528 & 65.34653465 & 63.63636364 &
\bfseries 74.36657682 & 47.5660066 & 53.63636364 \\

\midrule

$\text{Gemini}_5$ &
97 & 55 & 6 &
47.16981132 & 41.58415842 & 36.36363636 &
42.57861635 & 36.13861386 & 31.81818182 \\

$\text{ChatGPT}_5$ &
138 & 80 & 8 &
83.01886792 & 72.27722772 & 54.54545455 &
70.6918239 & 58.53960396 & 49.09090909 \\

$\textsc{CcfgT5}_{10}$ &
156 & 97 & 11 &
83.64779874 & \bfseries 76.23762376 & \bfseries 90.9090901 &
68.46960168 & \bfseries 65.32178218 & \bfseries 79.09090909 \\

\midrule
\midrule

Ground-truth &
159 & 101 & 11 &
100 & 100 & 100 &
81.50943396 & 85.66831683 & 90 \\

\bottomrule
\end{tabular}
\end{table*}

\subsection{Analysis of Experimental Results}

\tableref{tab:validity_effectiveness} presents statistics for test cases
generated by either baseline algorithms or CCFGs. CCFG-based test cases with
\ccfgtf[10] exhibit the highest set-based validity and both types of
effectiveness, achieving 81.18\%, 42.26\% and 67.73\%. In contrast, direct
generation using ChatGPT shows the highest element-based validity with 91.38\%.
Furthermore, when comparing the results of direct generation of ChatGPT and
those of CCFG-based ChatGPT$_5$, the latter shows higher effectiveness~(65.29\%
compared to 63.54\%).

It is important to note that the differences between the element-based validity
and the set-based validity of CCFG-based approaches are relatively smaller than
those in other categories. This suggests that the validity of each test case
generated by the CCFG-based approach is more consistent for each problem
specification. Once correct grammar and constraints are established, only valid
test cases are generated. Additionally, when considering ChatGPT, the set-based
validity over well-defined test sets with the CCFG-based approach is
93.36\%~($77.86\% \times 271 / 226$), which is higher than the validity of
direct generation at 78.97\%. This indicates that the CCFG-based approach is
more likely to fail in generating test cases than producing invalid ones.

As mentioned in \sectref{sec:dataset}, we use a human-labeled grammar as the
ground-truth only when the grammar can parse all the public and private test
cases in the CodeContests dataset. Therefore, the validity of these test cases
is represented by the ratio of problems with well-defined test case sets to the
number of total problems: 76.75\%~(208 out of 271) and 99.63\%~(270 out of 271).
We also anticipate that the element-based and set-based effectiveness could
reach $52.51\%$ and $83.40\%$, respectively, which are of ground-truth grammars,
if the model successfully generates semantically correct grammars.

\begin{tcolorbox}[boxsep=1pt,left=4pt,right=4pt,top=4pt,bottom=4pt]
\textbf{Observation 1:}~CCFGs are especially effective for problems with more
complex input specifications.
\end{tcolorbox}

\tableref{tab:difficulty} shows the relationship between the difficulty of test
case specifications, set-based validity, and effectiveness. We observe that the
direct test case generation using ChatGPT achieves the highest validity and
effectiveness, with $88.68\%$ and $74.37\%$ for problems with easy
specifications. In contrast, \ccfgtf[10] shows the highest results for problems
with normal or hard specifications.

Interestingly, \ccfgtf[10] exhibits a less pronounced tendency for validity to
decrease as difficulty increases compared to other methods. This observation
also highlights the robustness of \ccfgtf\ in understanding complex
specifications.

\begin{table*}[ht!]
\caption{
    Set-based effectiveness with respect to the length of test cases. For the
    columns of~S+M, S+L, and M+L, we used the union of two sets selected from
    the of short~(S), medium~(M) and long~(L) test cases. Note that the fifth
    column~(Mixed) uses the same sets of test cases as in
    \tableref{tab:validity_effectiveness}, which consists of 4~short, 3~medium
    and 3~long test cases.
}
\label{tab:length}
\centering

\sisetup{round-mode=places,round-precision=2,detect-weight=true}
\begin{tabular}{l SSSS SSS S}
\toprule
\multirow{2}{*}[-2pt]{\bf Method} &
\multicolumn{4}{c}{\textbf{Set-based effectiveness w/ 10~test cases}} & 
\multicolumn{3}{c}{\textbf{w/ 20~test cases}} & 
{\textbf{w/ 30~test cases}} \\

\cmidrule(lr){2-5}
\cmidrule(lr){6-8}
\cmidrule(lr){9-9}

&
{Short} & {Medium} & {Long} & {Mixed} &
{S+M}  & {S+L} & {M+L} &
{Short+Medium+Long} \\

\midrule

$\text{Gemini}_5$ &
32.76752768 & 34.760147601476 & 36.34686347 &  39.74169742 &
37.63837638376384 &  40.6642066420664 & 39.3726937269372 &
41.66051660516604 \\

$\text{ChatGPT}_5$ &
53.50553506 & 57.25092251 & 60.34440344 &  65.28597786 &
62.29397293972939 &  67.6568265682657 & 65.3813038130381 &
68.61623616236162 \\

$\textsc{CcfgT5}_{10}$ &
57.85670357 & 61.04551046 & 64.0498155 & \bfseries 67.72755228 &
65.9840098400984 & \bfseries 71.2669126691267 & 68.330258302583 &
\bfseries 72.0418204182041 \\

\midrule

Ground-truth &
71.06396064 & 73.13653137 & 78.20110701 &  83.40405904 &
80.60578105781057 &  86.7527675276752 & 83.330258302583 &
88.11808118081181 \\

\bottomrule
\end{tabular}
\end{table*}

\begin{tcolorbox}[boxsep=1pt,left=4pt,right=4pt,top=4pt,bottom=4pt]
\textbf{Observation 2:}~Longer test cases tend to be more
(element-wise)~effective, but the set of test cases with various lengths is the
most effective.
\end{tcolorbox}

\tableref{tab:length} presents the relationship between the length of test cases
and effectiveness. This result indicates that the variation in intervals during
generation impacts the efficiency of the resulting test case set, suggesting
there is potential for increasing efficiency by improving the generation
algorithm of CCFG. We also observe that the union of short and long test cases
exhibits the highest effectiveness among combinations from the results using
20~test cases. This indicates that short and long test cases identify different
incorrect problems.

Consequently, we conclude that utilizing a CCFG-based approach not only
eliminates the need for individual test case validation but also enhances the
effectiveness of test cases, particularly when input specifications are complex.
Additionally, there is potential to further increase effectiveness by generating
more test cases, which is straightforward with CCFGs, along with improved
heuristics for diverse generation.

\begin{tcolorbox}[boxsep=1pt,left=4pt,right=4pt,top=4pt,bottom=4pt]
\textbf{Observation 3:}~\ccfgtf\ generates more valid and general grammars than
LLMs such as ChatGPT and Gemini.
\end{tcolorbox}

\begin{table*}
\caption{
    Generality and semantic equality of generated grammars to ground-truth
    grammars. A generated grammar is semantically equivalent to the ground-truth
    grammar if and only if the grammar is valid and general.
}
\label{tab:generality_semantic_equivalence}

\sisetup{round-mode=places,round-precision=2,detect-weight=true}
\begin{tabular}{lSSSSS}
\toprule
\multirow{2}{*}[-2pt]{\bf Method} &
\multicolumn{2}{c}{\textbf{Generality} (\%)} &
{\textbf{Validity} (\%)} &
\multicolumn{2}{c}{\textbf{Equality} (\%)} \\
\cmidrule(lr){2-3}
\cmidrule(lr){4-4}
\cmidrule(lr){5-6}
&
{Element-based} & {Set-based} & {Set-based} &
{Semantic} & {Syntactic} \\

\midrule

Gemini$_1$ &
14.1768837 & 13.28413284 & 15.49815498 &
12.91512915 & 3.6900369 \\
 
Gemini$_5$ &
42.84013808 & 41.69741697 & 44.64944649 &
40.95940959 & 16.236162361623617 \\
 
\addlinespace[0.2em]

ChatGPT$_1$ &
59.87382454 & 55.7195572 & 66.42066421 &
54.61254613 & 14.391143911439114 \\
 
ChatGPT$_5$ &
72.93179383 & 71.21771218 & 77.8597786 &
70.11070111 & 29.88929889 \\

\addlinespace[0.2em]

\ccfgtf[1] &
19.95000595 & 19.55719557 & 18.81918819 &
18.08118081 & 11.07011070110701 \\
 
\ccfgtf[10] & \bfseries
83.44244733 & \bfseries 81.18081181 &  \bfseries 81.18081181 &  \bfseries
78.22878229 &  \bfseries 49.815498154981555 \\

\bottomrule
\end{tabular}
\centering
\end{table*}

The evaluation results presented in
\tableref{tab:generality_semantic_equivalence} demonstrate \ccfgtf\ produces the
most general grammars~(81.18\%) and those that are semantically
equivalent~(78.23\%). Notably, the average element-based generality closely
aligns with the average set-based generality across all methods, with
differences of less than or equal to 4.15\%p for ChatGPT$_5$. Since this
difference reflects the ratio of ground-truth test cases parsed by non-general
grammars, it suggests that determining whether a grammar is general using only a
few sampled ground-truth test cases is reliable. When we compare \ccfgtf\ and
LLMs, the ratios of syntactically different grammars to semantically equivalent
grammars are 42.63\%~($1 - 16.24/40.96$) and 63.68\%~($1 - 49.82/78.23$) for GPT
and \ccfgtf, respectively. It indicates that LLMs generate more syntactically
variant grammars compared to the fine-tuned CodeT5 model. Additionally, we use
Jaccard similarity~$J(G, V)$, which represents the ratio of the intersection
to the union to measure the similarity between the set~$G$ of general grammars
and the set of valid grammars~$V$:
\begin{align*}
    J(G, V)
    &= \frac{\text{(Semantic Equality)} \times 100 (\%)}
    {\text{(Set-General.)} + \text{(Set-Valid.)} - \text{(Sem. Equ.)}}.
\end{align*}
As a result, all similarities are greater than 80\%, with \ccfgtf[10] showing
the largest similarity with 92.98\%. This indicates that the most valid grammars
are also general, and they can generate all the test cases based on how we
sample the variables.

\subsection{Case Study}

We analyzed several failure cases where our proposed \ccfgtf~model struggles to
generate accurate CCFGs and their associated constraints. Frequently, the model
misinterprets natural language constraints, converting them incorrectly into
numerical constraints, and often handles grammars in natural language format
rather than structured CCFG. This leads to incomplete coverage of specified
constraints and the omission of additional implicit constraints. Also, the model
fails to cover all the constraints present in the CCFG, along with some unknown
constraints that are not included in the problem specification.

In \exampleref{example:failure-1}, the model extracts a constraint for a
variable~$n$ in natural language rather than converting it in to the expected
formal expression~$n \le 10^5$. Note that every constraint in ground-truth
grammars consists solely of formal expressions.

Our model encounters difficulties in understanding the necessity of translating
the natural language explanation into a formal representation. This failure is
caused by the fact that most specifications in our training dataset contain
explicit formal representations for constraints, while this example has such
constraints in natural language.

\begin{example}[139\_D. Digit Permutations from Codeforces]
\rm
\label{example:failure-1}
\hfill
\begin{center}
\begin{minipage}{0.95\linewidth}
\begin{Verbatim}[breaklines,frame=single,fontsize=\small,breaksymbolleft=]
The first line contains a positive integer n --- the original number. The number of digits in this number does not exceed 10^5. The number is written without any leading zeroes.
\end{Verbatim}
\vspace{2pt}
\begin{Verbatim}[breaklines,frame=single,fontsize=\small,breaksymbolleft=]
Grammar: ["<S> -> n"]
Constraints: ["Thenumberdoesnotexceed10^5."]
\end{Verbatim}
\end{minipage}
\end{center}
\end{example}

Additionally, in \exampleref{example:failure-2}, \ccfgtf\ erroneously conflates
\texttt{xi} as a variable that does not use a counter. This indicates that the
model fails to understand the meaning of ``\texttt{Each line}'', which implies a
use of a counter for \texttt{x}.

In contrast, the model accurately produces \texttt{x\_i} as a variable that uses
counters in \texttt{Constraint}. This inconsistency arises because we compose
\ccfgtf\ with two independent fine-tuned CodeT5~modules for grammars and
constraints, which assign different meanings for \texttt{xi}.

\begin{example}[103\_C. Russian Roulette from Codeforces]
\rm
\label{example:failure-2}
\hfill
\begin{center}
\begin{minipage}{0.95\linewidth}
\begin{Verbatim}[breaklines,frame=single,fontsize=\small,breaksymbolleft=]
The first line contains three integers n, k and p (...) Then follow p lines; they are the queries. Each line contains one integer xi (1 <= xi <= n) (...)
\end{Verbatim}
\vspace{2pt}
\begin{Verbatim}[breaklines,frame=single,fontsize=\small,breaksymbolleft=]
Grammar: [
  ...,
  "<Y_i> -> <Y_i-1> <n> xi",
  "<Y_1> -> xi"
]
Constraints: [..., "1<=x_i<=n"]
\end{Verbatim}
\end{minipage}
\end{center}
\end{example}

These failures are caused by uncommon natural language expressions for
constraints and inconsistent variable representations in the CodeContests
dataset. Data augmentation and normalization can address these issues.

\section{Conclusions}

We presented a novel framework leveraging CCFGs for ATCG in competitive
programming. By translating input specifications into formal grammars, our
method bridges the gap between specification complexity and test case validity,
offering substantial improvements in the accuracy and coverage of generated test
cases. Experiments highlight the effectiveness of our approach in distinguishing
incorrect algorithms and ensuring specification compliance.

We will expand CCFGs to handle broader input domains and optimizing sampling
strategies. By advancing these directions, our methodology could serve as a
foundation for scalable, reliable test case generation across diverse software
engineering applications.

\subsection*{Acknowledgment}

Sung and Han were supported by the NRF grant~(RS-2025-00562134) and the AI
Graduate School Program at Yonsei University~(RS-2020-II201361) funded by the
Korean government~(MSIT). Aditi and Ko were supported by the NRF
grant~(RS-2023-00208094) funded by MSIT.

\bibliographystyle{named}
\bibliography{bibliography}

\end{document}